\documentstyle[subeqn,epsfig]{elsart}
\newcommand{\vect}[1]{{\rm {\bf #1}}}
\newcommand{\rmt}[1]{\tiny\rm #1}
\newcommand{\lesim}{\raisebox{-0.55ex}{$\stackrel{\displaystyle <}{\sim}$}}

\begin{document}
\begin{frontmatter}

%
\title{Possible formulations for three-charged particles correlations in 
terms of Coulomb wave functions}

\author[Toba]{T. Mizoguchi\thanksref{takuya},}
\author[Shinshu]{M. Biyajima\thanksref{minoru}}
\address[Toba]{Toba National College of Maritime Technology, Toba 517-8501, 
Japan}
\address[Shinshu]{Department of Physics, Faculty of Science, Shinshu 
University, Matsumoto 390-8621, Japan}
 
\thanks[takuya]{E-mail: mizoguti@toba-cmt.ac.jp}
\thanks[minoru]{E-mail: biyajima@azusa.shinshu-u.ac.jp}

%
\begin{abstract}
The recent data for Bose-Einstein Correlations (BEC) of three-charged 
particles obtained by NA44 Collaboration have been analysed using theoretical 
formula with Coulomb wave functions. It has been recently proposed by Alt et 
al. It turns out that there are discrepancies between these data and the 
respective theoretical values. To resolve this problem we seek a possibly 
modified theoretical formulation of this problem by introducing the degree of 
coherence for the exchange effect due to the BEC between two-identical bosons. 
As a result we obtain a modified formulation for the BEC of three-charged 
particles showing good agreement with the data. Moreover, we investigate 
physical connection between our modified formulation and the core-halo model 
proposed by Cs\"org\H o et al. Our study indicates that the interaction region 
estimated by the BEC of three-charged particles in the S + Pb collisions at 
200 GeV/c per nucleon is equal to about 1.5 fm$\sim$1.8 fm.
\end{abstract}
\end{frontmatter}

%
\section{Introduction}
One of the most interesting subjects in high energy heavy-ion collisions is 
study of the higher order Bose-Einstein Correlation (BEC) effect 
\cite{biya90,cram91,hein97,heis97,alt99,huma99} (known also as the HBT or the 
GGLP effect, or as the hadron interferometry 
\cite{hanb56,gold59,gyul79,boal90}). From data on BEC we can (in principle) 
infer the size of the interaction region and therefore estimate the energy 
densities reached in high energy collisions. Such work is a necessary task in 
the search for the quark-gluon plasma \cite{bert89,weid99} - a new, 
hypothetical form of matter. 

To get more precise sizes of the interaction regions, we have to take into 
account the final state interactions among the charged particles 
\cite{prat86,biya94}. A great advance in this direction for the BEC of the 
three-charged particles has been recently made by Alt et. al. \cite{alt99}. 
They have derived a correction formula for the raw data introducing 
distribution functions of the charged particles. Their formulation is based on 
the plane wave functions and on the Coulomb wave functions, assuming that 
produced hadrons are already in the asymptotic region of the Coulomb 
interactions where the strong interaction already vanishes 
\cite{merk78,brau89}. It amounts in the following correction factor $K_{Coul}$ 
due to the Coulomb effect for identical three-charged particles:
\footnote{
The correlation functions for two and three-identical particles are given as 
usual by 
$$
\frac{N^{(2+\ {\rm or }\ 2-)}}{N^{BG}} = \frac{P(\vect k_1,\ \vect k_2)}
{P(\vect k_1)P(\vect k_2)} \quad{\rm and}\quad 
\frac{N^{(3+\ {\rm or }\ 3-)}}{N^{BG}} = \frac{P(\vect k_1,\ \vect k_2,\ 
\vect k_3)}{P(\vect k_1)P(\vect k_2)P(\vect k_3)}\ ,
$$
where $\vect k_i$ is the momentum of particle $i$, and $P(\vect k_1,\ 
\vect k_2)$ and $P(\vect k_1,\ \vect k_2,\ \vect k_3)$ are two and three 
particles probability densities, respectively. The probability densities for 
two-identical particles case can be written as, 
$$
\int \int \left|\psi_{\vect k_1\vect k_2}^{\rm BE}(\vect x_1,\ 
\vect x_2)\right|^2\rho (\vect x_1)\rho (\vect x_2) d^3 \vect x_1d^3 
\vect x_2\ ,
$$
where
$$
\psi_{\vect k_1\vect k_2}^{\rm BE}(\vect x_1,\ \vect x_2) = 
\frac 1{\sqrt 2}[\psi_{\vect k_1\vect k_2}(\vect x_1,\ \vect x_2) + 
\psi_{\vect k_1\vect k_2}(\vect x_2,\ \vect x_1)]\ ,
$$
$\rho (\vect x_i)$ stand for the source functions of particle $i$.
} 
\begin{equation}
  K_{Coul} = \frac{N_{Coul}}{D_{plane}}\ .
  \label{eq1}
\end{equation}
The denominator $D_{plane}$ is given by ($\rho (\vect x_i)$ are distribution 
functions of charged particles):
\begin{eqnarray}
  D_{plane} & \cong & \frac 16 \int 
  d^3 \vect x_1\rho (\vect x_1) 
  d^3 \vect x_2\rho (\vect x_2) 
  d^3 \vect x_3\rho (\vect x_3)\nonumber\\
  &&\quad \cdot \left|
    e^{i(\vect k_1\cdot \vect x_1+\vect k_2\cdot \vect x_2+\vect k_3\cdot 
    \vect x_3)}
  + e^{i(\vect k_1\cdot \vect x_2+\vect k_2\cdot \vect x_1+\vect k_3\cdot 
  \vect x_3)}\right .\nonumber\\
  &&\quad + 
    e^{i(\vect k_1\cdot \vect x_2+\vect k_2\cdot \vect x_3+\vect k_3\cdot 
    \vect x_1)}
  + e^{i(\vect k_1\cdot \vect x_1+\vect k_2\cdot \vect x_3+\vect k_3\cdot 
  \vect x_2)}\nonumber\\
  &&\quad \left . + e^{i(\vect k_1\cdot \vect x_3+\vect k_2\cdot \vect x_1 +
  \vect k_3\cdot \vect x_2)}
  + e^{i(\vect k_1\cdot \vect x_3+\vect k_2\cdot \vect x_2+\vect k_3\cdot 
  \vect x_1)}\right|^2\ ,
  \label{eq2}
\end{eqnarray}
The numerator $N_{Coul}$ has the following form:
\begin{eqnarray}
  N_{Coul} & \cong & \frac 16 \int 
  d^3 \vect x_1\rho (\vect x_1) 
  d^3 \vect x_2\rho (\vect x_2) 
  d^3 \vect x_3\rho (\vect x_3)\nonumber\\
  &&\quad \cdot \left|
  \psi_{\vect k_1\vect k_2}^C(\vect x_1,\ \vect x_2) 
  \psi_{\vect k_2\vect k_3}^C(\vect x_2,\ \vect x_3) 
  \psi_{\vect k_3\vect k_1}^C(\vect x_3,\ \vect x_1)\right .\nonumber\\
  &&\quad + 
  \psi_{\vect k_1\vect k_2}^C(\vect x_1,\ \vect x_3) 
  \psi_{\vect k_2\vect k_3}^C(\vect x_3,\ \vect x_2) 
  \psi_{\vect k_3\vect k_1}^C(\vect x_2,\ \vect x_1)\nonumber\\
  &&\quad + 
  \psi_{\vect k_1\vect k_2}^C(\vect x_2,\ \vect x_1) 
  \psi_{\vect k_2\vect k_3}^C(\vect x_1,\ \vect x_3) 
  \psi_{\vect k_3\vect k_1}^C(\vect x_3,\ \vect x_2)\nonumber\\
  &&\quad + 
  \psi_{\vect k_1\vect k_2}^C(\vect x_2,\ \vect x_3) 
  \psi_{\vect k_2\vect k_3}^C(\vect x_3,\ \vect x_1) 
  \psi_{\vect k_3\vect k_1}^C(\vect x_1,\ \vect x_2)\nonumber\\
  &&\quad + 
  \psi_{\vect k_1\vect k_2}^C(\vect x_3,\ \vect x_1) 
  \psi_{\vect k_2\vect k_3}^C(\vect x_1,\ \vect x_2) 
  \psi_{\vect k_3\vect k_1}^C(\vect x_2,\ \vect x_3)\nonumber\\
  &&\quad \left . + 
  \psi_{\vect k_1\vect k_2}^C(\vect x_3,\ \vect x_2) 
  \psi_{\vect k_2\vect k_3}^C(\vect x_2,\ \vect x_1) 
  \psi_{\vect k_3\vect k_1}^C(\vect x_1,\ \vect x_3)\right|^2\ .
  \label{eq3}
\end{eqnarray}
Here $\psi_{\vect k_i\vect k_j}^C(\vect x_i,\ \vect x_j)$ are the Coulomb wave 
functions of the respective 2-body collision expressed as, 
\begin{equation}
  \psi_{\vect k_i \vect k_j}^C(\vect x_i,\ \vect x_j) =  \Gamma(1 + i\eta_{ij})
  e^{-\pi \eta_{ij}/2} e^{ i \vect k_{ij} \cdot \vect{r}_{ij} }
  F[- i \eta_{ij},\ 1;\ i ( k_{ij} r_{ij} - \vect k_{ij} \cdot \vect r_{ij} )],
  \label{eq4}
\end{equation}
with $\vect r_{ij} = (\vect x_i - \vect x_j)$, $\vect k_{ij} = (\vect k_i - 
\vect k_j)/2$, $r_{ij} = |\vect r_{ij}|$, $k_{ij} = |\vect k_{ij}|$ and 
$\eta_{ij} = m\alpha/k_{ij}$. $F[a,\ b;\ x]$ and $\Gamma(x)$ 
are the confluent hypergeometric function and the Gamma function, 
respectively. In order to use Eqs. (\ref{eq1}), (\ref{eq2}) and (\ref{eq3}), 
one has to assume first some shapes and sizes for the source functions. In 
fact, this is the procedure already used in Ref. \cite{na4499} by NA44 
Collaboration: 
$$
{\rm Corrected\ data} = ({\rm raw\ data}) \times K_{\rm spc} \times 
K_{\rm acceptance} \times K_{Coul}\ ,
$$
where $K_{spc}$ and $K_{acceptance}$ denote the effect of multiparticle 
production in the single particle spectra and the acceptance effect in the 
experiment.

In this paper, we would like to adopt a different point of view for 
Eq. (\ref{eq3}). As is seen in Ref. \cite{biya94}, the BEC of identical 
two-charged pions can also be analysed by the Coulomb wave functions. It is 
therefore reasonable to expect that the numerator $N_{Coul}$ is the main 
theoretical ingredient in analysis of the BEC of three-charged particles. We 
argue therefore that
\begin{equation}
  N^{(3+\ {\rm or }\ 3-)}/N^{BG} \equiv \mbox{$C$}\times N_{Coul}\ ,
  \label{eq5}
\end{equation}
where we have introduced the normalization factor $C$, which corresponds to 
the asymptotic value of the BEC. Using Eq. (\ref{eq5}) we can now (with the 
help of the CERN-MINUIT program) analyse data of Ref. \cite{na4499} using 
Gaussian source distributions of radii $R$, $\rho(\vect x)= 
\frac{1}{(2\pi R^2)^{3/2}} \exp\left[-\frac{{\vect x}^2}{2R^2}\right]$
\footnote{
It should be remembered that NA44 Collaboration data are for the variable 
$$
Q_3^2(4D) = (k_1 - k_2)^2 + (k_2 - k_3)^2 + (k_3 - k_1)^2 
$$
where $k_i$ are four-momentum of charged particles. $Q_3 = \sqrt{Q_3^2}$. 
However, in our calculations we assume that $q_{0,ij}^2 = ({\rm k}_{0i} - 
{\rm k}_{0j})^2 \approx 0$ and use, instead,
$$
Q_3^2(3D) = (\vect k_1 - \vect k_2)^2 + (\vect k_2 - \vect k_3)^2 + 
(\vect k_3 - \vect k_1)^2\ .
$$ 
There is an approximation between them
$$
Q_3(4D) \cong Q_3(3D) - \sum q_{0,ij}^2/2Q_3(3D)\ .
$$
$Q_3(4D)$ depends on sum of squared energy differences. 
}. 

In the next paragraph, we analyse the data of NA44 Collaboration \cite{na4499} 
by Eq. (\ref{eq5}). In the third paragraph we shall derive a modefied 
theoretical formula for 3-particle BEC introducing the degree of coherence 
parameter into Eq. (\ref{eq5}). This formula will be then used in the 4th 
paragraph for the re-analyses of the experimental data \cite{na4499}. In the 
5th paragraph, we investigate whether or not there is physical connection 
between our study and the core-halo model \cite{csorgo99}. Concluding remarks 
are given in the final paragraph. 

%
\section{Application of Eq. (\ref{eq5}) to the data by NA44 Collaboration}
\begin{figure}[htb]
  \centering
  \epsfig{file=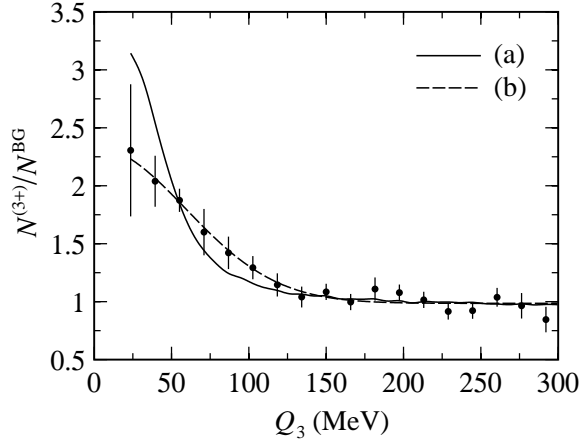,height=60mm}
  \caption{Analysis of 3$\pi^+$ BEC in S + Pb collision \cite{na4499}. (a) and 
  (b) are results of Eqs. (\ref{eq5}) and (\ref{eq6}), respectively. 
The error bars are sum of statistical and systematic errors.}
  \label{fig1}
\end{figure}

Here we analyse the data by Eq. (\ref{eq5}). As can be seen in Fig. \ref{fig1} 
and Table \ref{table1}, there are some discrepancies between the data points 
and theoretical values calculated by means of Eq. (\ref{eq5}). The minimum 
$\chi^2$ value, 17.6 in Table \ref{table1}, seems to be big, as the number of 
the data points are considered. Thus we would like to know why this equation 
cannot explain the data \cite{na4499}. We know that there are several possible 
reasons due to effects of the partial coherence, the contamination 
\cite{cramer96} and the long-lived resonances \cite{csorgo99}. At present, we 
consider the effect of the possible partial coherent of produced pions. In 
fact, authors of Ref. \cite{na4499} have used not the equivalence of Eq. 
(\ref{eq5}) but the following formula instead (cf., Ref. \cite{gyul79,deut78}):
\begin{equation}
  \frac{N^{(3+)}}{N^{BG}} = C\left(1+\lambda_3 e^{-R_3^2Q_3^2}\right)\ .
  \label{eq6}
\end{equation}
It contains one more parameter, $\lambda_3$, which can be regarded as a kind 
of effective degree of coherence and which, in our opinion, should therefore 
occur also somehow in Eq.\ (\ref{eq5}). 

\vspace{5mm}
\begin{table}[htb]
  \centering
  \caption{Estimated values for the data \cite{na4499} by Eqs. (\ref{eq5}) and 
  (\ref{eq6}) using CERN-MINUIT program.}
  \label{table1}
  \begin{tabular}{ccccc}
  \hline
  Formulas & $C$ & $R$ [fm] & $\lambda_3$ &$\chi^2/N_{dof}$\\
  \hline
  Eq. (\ref{eq5}) & 0.941$\pm$0.026 & 2.47$\pm$0.14 & --- & 17.6/16\\
  Eq. (\ref{eq6}) & 0.986$\pm$0.028 & 2.36$\pm$0.26 & 1.37$\pm$0.19 & 7.8/15\\
  \hline
  \end{tabular}
\end{table}

%
\section{Diagram Decomposition of Eq. (\ref{eq5})}
First of all, we have to find a possible way for the introduction of the 
degree of coherence parameter $\lambda$ into Eq. (\ref{eq5}). Let us therefore 
examine the plane wave (PW) approximations of the Coulomb wave 
functions:
\begin{subeqnarray}
  \label{eq7}
  A(1) &=& 
  \psi_{\vect k_1\vect k_2}^C(\vect x_1,\ \vect x_2)
  \psi_{\vect k_2\vect k_3}^C(\vect x_2,\ \vect x_3)
  \psi_{\vect k_3\vect k_1}^C(\vect x_3,\ \vect x_1)\nonumber\\
  && \stackrel{\rmt{PW}}{\longrightarrow}
  e^{ i \vect k_{12} \cdot \vect r_{12}}
  e^{ i \vect k_{23} \cdot \vect r_{23}}
  e^{ i \vect k_{31} \cdot \vect r_{31}}
  = e^{ (3/2)i (\vect k_1 \cdot \vect x_1
          + \vect k_2 \cdot \vect x_2
          + \vect k_3 \cdot \vect x_3)}\ ,\\
A(2) &=& 
  \psi_{\vect k_1\vect k_2}^C(\vect x_1,\ \vect x_3) 
  \psi_{\vect k_2\vect k_3}^C(\vect x_3,\ \vect x_2) 
  \psi_{\vect k_3\vect k_1}^C(\vect x_2,\ \vect x_1)\nonumber\\
  && \stackrel{\rmt{PW}}{\longrightarrow} 
  e^{ i \vect k_{12} \cdot \vect r_{13}} 
  e^{ i \vect k_{23} \cdot \vect r_{32}} 
  e^{ i \vect k_{31} \cdot \vect r_{21}}
  = e^{ (3/2)i (\vect k_1 \cdot \vect x_1
          + \vect k_2 \cdot \vect x_3
          + \vect k_3 \cdot \vect x_2)}\ ,\\
  A(3) &=& 
  \psi_{\vect k_1\vect k_2}^C(\vect x_2,\ \vect x_1)
  \psi_{\vect k_2\vect k_3}^C(\vect x_1,\ \vect x_3)
  \psi_{\vect k_3\vect k_1}^C(\vect x_3,\ \vect x_2)\nonumber\\
  && \stackrel{\rmt{PW}}{\longrightarrow}
  e^{ i \vect k_{12} \cdot \vect r_{21}}
  e^{ i \vect k_{23} \cdot \vect r_{13}}
  e^{ i \vect k_{31} \cdot \vect r_{32}}
  = e^{ (3/2)i (\vect k_1 \cdot \vect x_2
          + \vect k_2 \cdot \vect x_1
          + \vect k_3 \cdot \vect x_3)}\ ,\\
  A(4) &=& 
  \psi_{\vect k_1\vect k_2}^C(\vect x_2,\ \vect x_3) 
  \psi_{\vect k_2\vect k_3}^C(\vect x_3,\ \vect x_1) 
  \psi_{\vect k_3\vect k_1}^C(\vect x_1,\ \vect x_2)\nonumber\\
  && \stackrel{\rmt{PW}}{\longrightarrow}
  e^{ i \vect k_{12} \cdot \vect r_{23}}
  e^{ i \vect k_{23} \cdot \vect r_{31}}
  e^{ i \vect k_{31} \cdot \vect r_{12}}
  = e^{ (3/2)i (\vect k_1 \cdot \vect x_2
          + \vect k_2 \cdot \vect x_3
          + \vect k_3 \cdot \vect x_1)}\ ,\\
  A(5) &=& 
  \psi_{\vect k_1\vect k_2}^C(\vect x_3,\ \vect x_1)
  \psi_{\vect k_2\vect k_3}^C(\vect x_1,\ \vect x_2)
  \psi_{\vect k_3\vect k_1}^C(\vect x_2,\ \vect x_3)\nonumber\\
  && \stackrel{\rmt{PW}}{\longrightarrow}
  e^{ i \vect k_{12} \cdot \vect r_{31}}
  e^{ i \vect k_{23} \cdot \vect r_{12}}
  e^{ i \vect k_{31} \cdot \vect r_{23}}
  = e^{ (3/2)i (\vect k_1 \cdot \vect x_3
          + \vect k_2 \cdot \vect x_1
          + \vect k_3 \cdot \vect x_2)}\ ,\\
  A(6) &=& 
  \psi_{\vect k_1\vect k_2}^C(\vect x_3,\ \vect x_2) 
  \psi_{\vect k_2\vect k_3}^C(\vect x_2,\ \vect x_1) 
  \psi_{\vect k_3\vect k_1}^C(\vect x_1,\ \vect x_3)\nonumber\\
  && \stackrel{\rmt{PW}}{\longrightarrow} 
  e^{ i \vect k_{12} \cdot \vect r_{32}} 
  e^{ i \vect k_{23} \cdot \vect r_{21}} 
  e^{ i \vect k_{31} \cdot \vect r_{13}}
  = e^{ (3/2)i (\vect k_1 \cdot \vect x_3
          + \vect k_2 \cdot \vect x_2
          + \vect k_3 \cdot \vect x_1)}\ .
\end{subeqnarray}

\begin{figure}[htb]
  \centering
  \epsfig{file=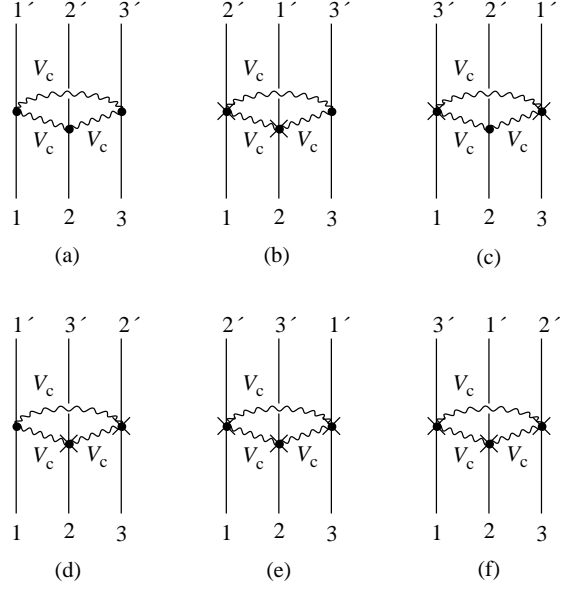,height=80mm}
  \caption{Diagram reflecting three-charged particles Bose-Einstein 
  Correlation  (BEC) and Coulombic potential ($V_c$). {\large $\times$} means 
  the exchange effect of BEC.}
  \label{fig2}
\end{figure}
Notice that, except for the factor $3/2$, exponential functions are the same 
expressions as those present in the integrand of Eq. (\ref{eq2}). This 
difference is attributed to the fact that Coulomb wave function used here 
describes two-charged particles collisions, therefore factor $3/2$ appears 
because there are relevant two-particle three combinations among three-charged 
particles.

Combining Eqs.\ (\ref{eq7}) and Figs.\ \ref{fig2}, we obtain the following 
three sets of equations:
\begin{subeqnarray}
\label{eq8}
  F_1 &=& \frac 16 \sum_{i=1}^6 A(i)A^*(i) 
\stackrel{\rmt{PW}}{\longrightarrow} 1\ ,\\
  F_2 &=& \frac 16 [ A(1)A^*(2) + A(1)A^*(3) + A(1)A^*(6) + A(2)A^*(4) + 
A(2)A^*(5)\nonumber\\
  && \quad + A(3)A^*(4) + A(3)A^*(5) + A(4)A^*(6) + A(5)A^*(6) + {\rm 
c.\ c.}]\nonumber\\
  && \quad \stackrel{\rmt{PW}}{\longrightarrow} 
  \mbox{ BEC between two-charged particles}\\
  && \quad\qquad \mbox{ (See Figs. \ref{fig2}\ (b)$\sim$(d))}\ 
,\nonumber\\
  F_3 &=& \frac 16 [ A(1)A^*(4) + A(1)A^*(5) + A(2)A^*(3) + A(2)A^*(6) + 
A(3)A^*(6)\nonumber\\
  && \quad + A(4)A^*(5) + {\rm c.\ c.}]\nonumber\\
  && \quad \stackrel{\rmt{PW}}{\longrightarrow} 
  \mbox{ BEC among three-charged particles}\\
  && \quad\qquad \mbox{ (See Figs. \ref{fig2}\ (e) and (f))}\ .\nonumber
\end{subeqnarray}
Combining now Eqs. (\ref{eq8}) and the concept of partial coherence for 
the BEC \cite{na4499,deut78}, we can introduce a coherence parameter 
$\sqrt{\lambda}$ for the single mark ({\large $\times$}) in Fig. \ref{fig2}. 
The $\lambda = 1$ corresponds to the totally chaotic source, which is the 
assumption behind Eq. (\ref{eq5}).
Taking into account the strength of the degree of coherence $\lambda$ between 
two-identical bosons and $\lambda^{3/2}$ among three-identical bosons in 
Figs. \ref{fig2}, we can finally express the BEC for three identical charged 
particles as:
\begin{equation}
  \frac{N^{(3+\ {\rm or }\ 3-)}}{N^{BG}} \cong C \int 
  d^3 \vect x_1\rho (\vect x_1) 
  d^3 \vect x_2\rho (\vect x_2) 
  d^3 \vect x_3\rho (\vect x_3)
  [F_1 + \lambda F_2 + \lambda^{3/2} F_3]\ .
  \label{eq9}
\end{equation}
Equation (\ref{eq9}) is the modified theoretical formula we were looking for. 
It differs from Eq. (\ref{eq5}) originally proposed by Alt et. al. in 
\cite{alt99} by the presence of the degree of coherence $\lambda$ and in the 
limit of $\eta_{ij} \to 0$ it becomes
\begin{equation}
  {\rm Eq.\ (\ref{eq9})} \stackrel{\eta_{ij} \to 0}{\longrightarrow} 
  C\left(1 + 3\lambda e^{-\frac 34 R^2Q_3^2} + 2\lambda^{3/2} 
  e^{-\frac 98 R^2Q_3^2}\right)\ ,
  \label{eq10}
\end{equation}
which is the extended formula proposed some time ago by Deutschmann et al. 
\cite{deut78}. 

It should be noticed that Eq.\ (\ref{eq9}) can be applied to data corrected 
only by the Gamow factor $G(\eta_{12})G(\eta_{23})G(\eta_{31})$ in an ideal 
case \cite{biya96}, because Eq.\ (\ref{eq9}) is described by the Coulomb wave 
functions including the Gamow factors (see Ref. \cite{liu86})
\footnote{
In other words, ideal data sets for Eq. (\ref{eq9}) are of the form of
$$
{\rm Corrected\ data} = ({\rm raw\ data}) \times K_{\rm spc} \times 
K_{\rm acceptance} \times K_{\rm Gamow}\ ,
$$
where $K_{\rm Gamow} = 1/(G(\eta_{12})G(\eta_{23})G(\eta_{31}))$.
}. 

%
\section{Reanalyses of NA44 Collaboration data by means of Eq.\ (\ref{eq9})}
\begin{figure}[htb]
  \centering
  \epsfig{file=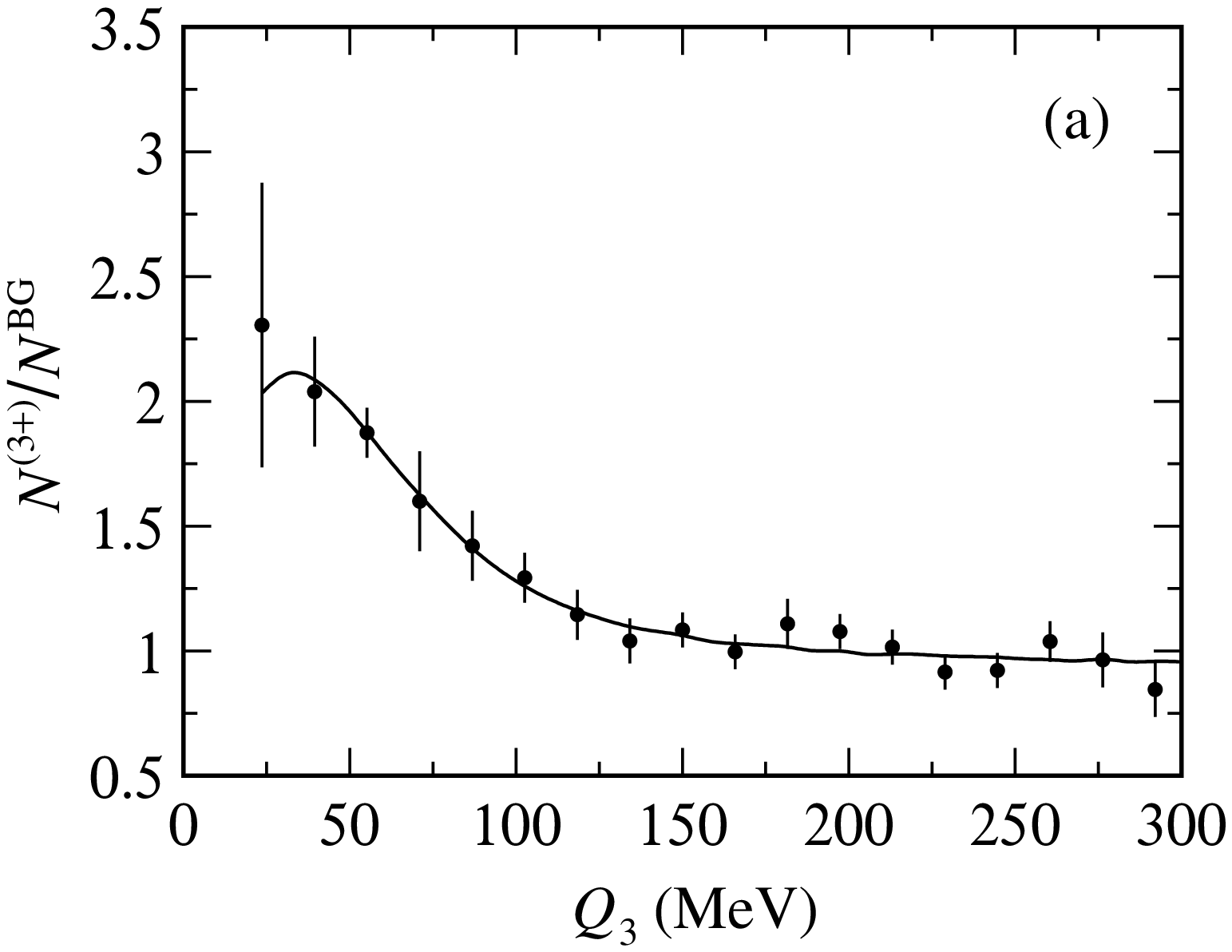,height=60mm}
  \epsfig{file=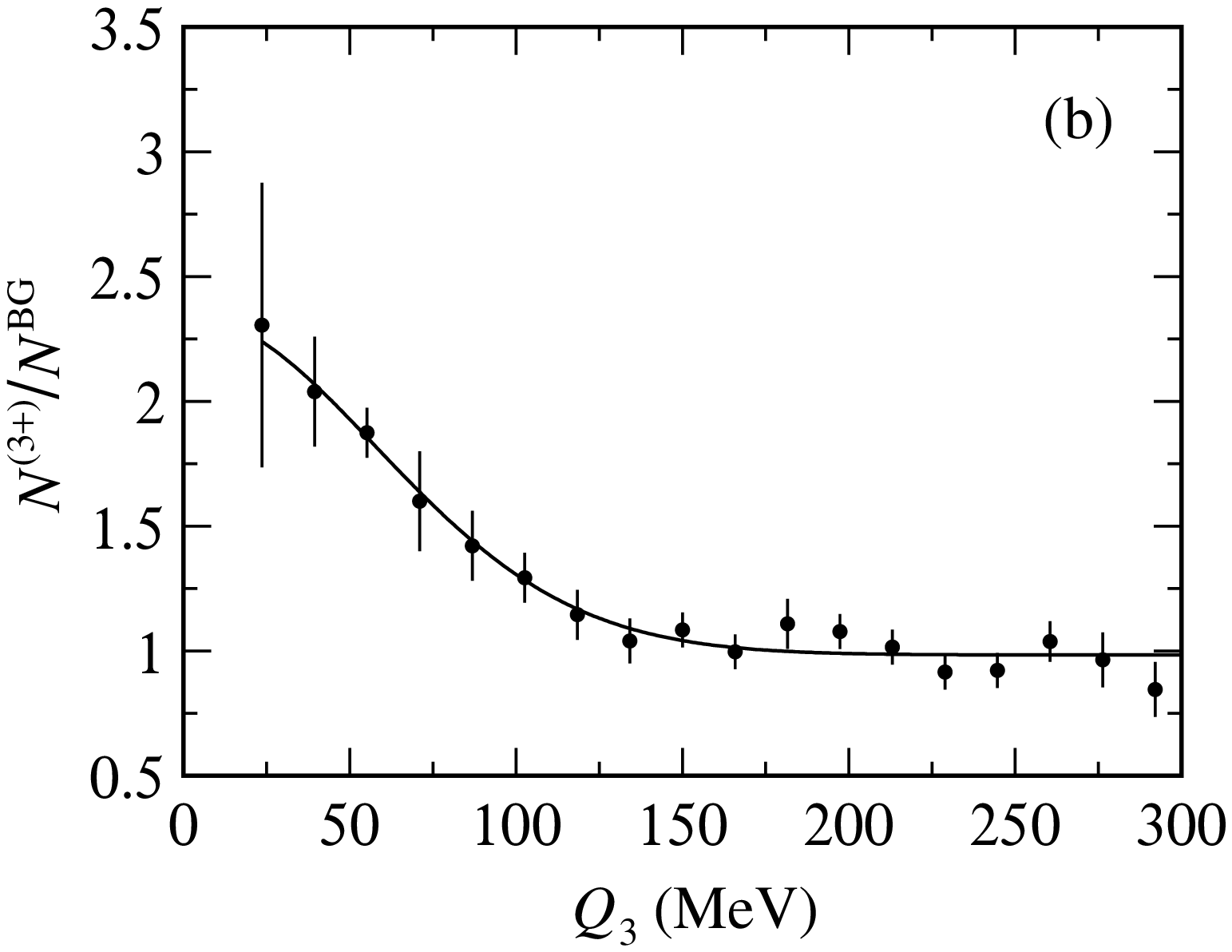,height=60mm}
  \caption{Reanalyses of 3$\pi^+$ BEC in S + Pb collision \cite{na4499}. 
  (a) is result of Eq. (\ref{eq9}). (b) is that of Eq. (\ref{eq10}).}
  \label{fig3}
\end{figure}

At present we have no data corrected only by the Gamow factors, therefore we 
apply Eq. (\ref{eq9}) to the analysis of NA44 Collaboration data \cite{na4499} 
using the CERN-MINUIT program. Our results are shown in Fig.\ \ref{fig3} and 
Table\ \ref{table2}. Comparing them  with those of Table\ \ref{table1}, it can 
be said that the $\chi^2$-value becomes smaller, i.e., the agreement is now 
better. The range of interaction becomes also smaller. For the sake of 
reference we present in Table \ref{table2} also results obtained by using 
Eq. (\ref{eq10}) .
\vspace{5mm}
\begin{table}[htb]
  \centering
  \caption{Reanalyses of 3$\pi^+$ BEC in S + Pb collision \cite{na4499} 
  by Eqs. (\ref{eq9}) and (\ref{eq10}).}
  \label{table2}
  \begin{tabular}{ccccc}
  \hline
  Formula & $C$ & $R$ [fm] & $\lambda$ &$\chi^2/N_{dof}$\\
  \hline
  Eq. (\ref{eq9}) & 0.917$\pm$0.032 & 1.53$\pm$0.20 & 0.55$\pm$0.07 & 6.7/15\\
  Eq. (\ref{eq10}) & 0.984$\pm$0.029 & 2.60$\pm$0.28 & 0.33$\pm$0.04 & 7.7/15\\
  \hline
  \end{tabular}
\end{table}

From results of Table \ref{table2}, we see that $R = 1.53$ fm by Eq. 
(\ref{eq9}) is small. The reason is attributed to the fraction of partially 
coherent effect ($\lambda$). From comparisons between results by Eqs. 
(\ref{eq9}) and (\ref{eq10}), it can be seen the interaction region becomes 
smaller, and the degree of coherence does conversely bigger due to the Coulomb 
interaction.

%
\section{Possible interpretation of $\lambda$ by core-halo model}
The core-halo model has been proposed by Cs\"org\H o et al. \cite{csorgo99}. 
We study whether or not there is physical connection between our previous 
formulation and theirs \cite{csorgo99}. To apply their model to Eq. 
(\ref{eq8}), first of all we use the following source functions
\begin{equation}
  \rho (x_i) = \rho_c (x_i) + \rho_{halo} (x_i)\ ,
  \label{eq11}
\end{equation}
where $\rho_c (x_i) = \frac{1}{(2\pi R^2)^{3/2}} 
\exp \left[-\frac{{\vect x_i}^2}{2R^2}\right]$ and $\rho_{halo} (x_i) = 
\frac{1}{(2\pi R_h^2)^{3/2}} \exp\left[-\frac{{\vect x_i}^2}{2R_h^2}\right]$. 
\begin{figure}[htb]
  \centering
  \epsfig{file=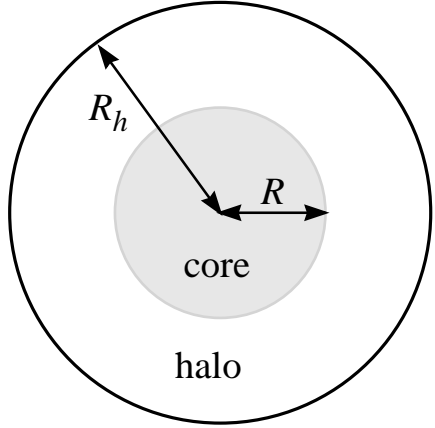,height=60mm}
  \caption{Physical picture of core-halo model.}
  \label{fig4}
\end{figure}

In concrete analyses, the radius of the halo becomes to be infinite, due to 
the effect of long-lived resonances ($R_h \to \infty$). The exchange function 
of the halo part due to the BEC, $E_{halo} = \exp(-Q^2R_h^2) \to 0$. Then we 
can interpret $\lambda$ in Eq. (\ref{eq9}) by their parameter, i. e., fraction 
of multiplicity from the core part, $f_c = \langle n_{core}\rangle/\langle 
n_{tot}\rangle$, as $\lambda = f_c^2$. In this case, it should be stressed 
that their parameter $p_c = \langle n_{co}\rangle/\langle n_{core}\rangle$, 
the fraction of the coherently produced multiplicity from the core part is 
zero.

Moreover, we can assume the laser optical approach for the core part 
\cite{csorgo99}, introducing a parameter $p = \langle n_{chao}\rangle/\langle 
n_{core}\rangle$ \cite{biya90}. Notice that $p = 1 - p_c$ \cite{csorgo99}. For 
the cross mark ({\large $\times$}) in Fig. \ref{fig2}, we assume two 
components of coherently ($1 - p$) and chaotically produced particles ($p$). 
Then we obtain the following expression
\begin{eqnarray}
  \frac{N^{(3+\ {\rm or }\ 3-)}}{N^{BG}} &\cong& C \left[\ \int 
  d^3 \vect x_1\rho_c (\vect x_1) 
  d^3 \vect x_2\rho_c (\vect x_2) 
  d^3 \vect x_3\rho_c (\vect x_3)
  (F_1 + f_c^2p^2 F_2 + f_c^3p^3 F_3)\right .\nonumber\\
  &&\qquad + \int 
  d^3 \vect x_1\rho_c (\vect x_1) \cdot \delta^3 (\vect x_1)
  d^3 \vect x_2\rho_c (\vect x_2) 
  d^3 \vect x_3\rho_c (\vect x_3) \nonumber\\
  &&\qquad\qquad \left . \cdot \left ( 2f_c^2p(1-p) F_2 + 3f_c^3p^2(1-p) F_3
  \right )\ \right ] \ .
  \label{eq12}
\end{eqnarray}
The effective degree of coherence, $\lambda_3^*$, the intercept at smallest 
$Q_3$ is given as
\begin{equation}
  \lambda_3^* = f_c^2(p^2+2p(1-p)) + f_c^3(p^3+3p^2(1-p))\ .
  \label{eq13}
\end{equation}
In Eq. (\ref{eq12}), as $p = 1$, we obtain Eq. (\ref{eq9}) with $\lambda = 
f_c^2$. On the contrary, as $f_c = 1$, we obtain an expression of laser 
optical approach \cite{biya90}. By making use of Eq. (\ref{eq12}) and an 
expression for the two-charged particles of the BEC, we can analyse the data 
by NA44 Collaboration. 

\begin{figure}[htb]
  \centering
  \epsfig{file=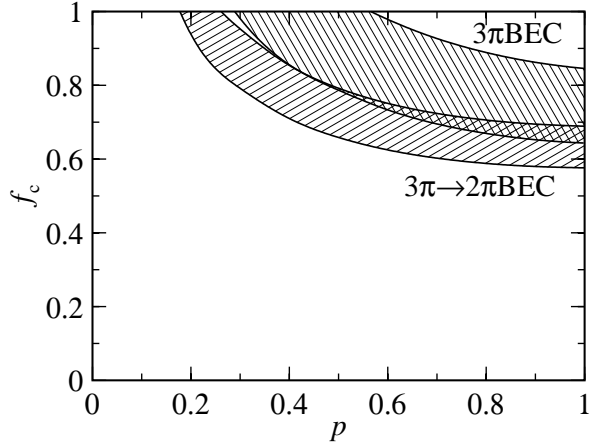,height=60mm}
  \caption{Sets of $f_c$ and $p$ estimated in analyses of BEC of two-charged 
  and three-charged pions using the CERN-MINUIT program. Widths of $f_c$'s 
  stand for error bar of $\pm 2\sigma$.}
  \label{fig5}
\end{figure}

For the two-charged particles of the BEC by means of the core-halo model with 
laser optical approach, we have the following equation,
\begin{eqnarray}
  \frac{N^{(2+\ {\rm or }\ 2-)}}{N^{BG}} &\cong& C \left[\int 
  d^3 \vect x_1\rho (\vect x_1) 
  d^3 \vect x_2\rho (\vect x_2) (G_1 + f_c^2p^2 G_2)\right .\nonumber\\
  &&\left . + \int 
  d^3 \vect x_1\rho (\vect x_1) 
  d^3 \vect x_2\rho (\vect x_2) \cdot \delta^3 (\vect x_2) 2f_c^2p(1-p)G_2
  \right ]\ ,
  \label{eq14}
\end{eqnarray}
where $G_1 = \frac 12 \left (\left |\psi_{\vect k_1 \vect k_2}^C(\vect x_1,\ 
\vect x_2)\right |^2  + \left |\psi_{\vect k_1 \vect k_2}^C(\vect x_2,\ 
\vect x_1)\right |^2\right )$ and $G_2 = {\rm Re}\left (\psi_{\vect k_1 
\vect k_2}^{C*}(\vect x_1,\ \vect x_2)\right .$ $\left .\cdot \psi_{\vect k_1 
\vect k_2}^C(\vect x_2,\ \vect x_1)\right )$. The effective degree of 
coherence $\lambda_2^*$ is given as
\begin{equation}
  \lambda_2^* = f_c^2(p^2+2p(1-p))\ .
  \label{eq15}
\end{equation}
Results of our analyses are shown in Table \ref{table3} and Figure \ref{fig5}. 
As is seen in them, the interaction ranges of $R$(core part) are estimated in 
the ranges of $1.5\ {\rm fm} < R{\rm (core)} < 1.8\ {\rm fm}$, and  $4.7\ 
{\rm fm} < R{\rm (core)} < 5.4\ {\rm fm}$, for the BEC of 3$\pi$ and $3\pi\to 
2\pi$, respectively.%
\begin{table}[htb]
  \centering
  \caption{Typical results from analyses of data of NA44 Collaboration by Eqs. 
  (\ref{eq12})$\sim$(\ref{eq15}).}
  \label{table3}
  \begin{tabular}{cccc}
  \hline
  \multicolumn{4}{c}{$3\pi$ BEC}\\
  \hline
  $p$ & 1.0 & 0.8 & 0.6\\
  \hline
  $f_c$ & $0.743\pm 0.050$ & $0.779\pm 0.055$ & $0.856\pm 0.061$\\
  $\lambda_3^*$ & 0.964 & 1.007 & 1.022\\
  $R$ (fm) & $1.53\pm 0.21$ & $1.72\pm 0.25$ & $1.87\pm 0.27$\\
  $\chi^2/N_{dof}$ & 6.7/15 & 6.6/15 & 6.6/15\\
  \hline \hline
  \multicolumn{4}{c}{$3\pi \to 2\pi$ BEC}\\
  \hline
  $p$ & 1.0 & 0.8 & 0.6\\ 
  \hline
  $f_c$ & $0.633\pm 0.028$ & $0.647\pm 0.029$ & $0.688\pm 0.032$\\
  $\lambda_2^*$ & 0.400 & 0.402 & 0.398\\
  $R$ (fm) & $4.69\pm 0.45$ & $5.34\pm 0.54$ & $5.85\pm 0.59$ \\
  $\chi^2/N_{dof}$ & 14.6/17 & 14.9/17 & 14.9/17\\
  \hline
  \end{tabular}
\end{table}
\vspace{5mm}

The common region between results of the BEC of three-pion and two-pion is 
roughly described by $0.6 \lesim p \lesim 1.0$ and $f_c \sim 0.7$.

%
\section{Concluding remarks}
We have derived the theoretical formula for the BEC of three-charged identical 
particle using both the Coulomb wave functions and the notion of the degree of 
coherence and compared it with the experimental data
\footnote{
For the numerical calculations of Eqs. (\ref{eq5}) and (\ref{eq9}) (in order 
to save the CPU-time), we have first calculated 10$^3$ k values of the Coulomb 
wave functions, which were then used together with some interpolation 
procedure during the concrete calculations. In this way we could make use of 
the CERN-MINUIT program in our analyses.
}. 
Historically the degree of coherence in the BEC of the two-identical bosons 
has been introduced by experimentalist \cite{deut78}, and theoretical works in 
this direction have been performed in Ref. \cite{gyul79}.

Our first analyses suggest that the degree of coherence $\lambda$ is a 
necessary ingredient also for the BEC of three-charged particles, in the same 
way as it was for the BEC for two-charged identical particles. This fact means 
that the source producing finally observed particles is not purely chaotic. 
However the investigation in 5th paragraph elucidates that our modified 
formulation can be interpreted by the core-halo model. Our degree of coherence 
$\lambda$ is equal to $f_c^2$ [introduced in Ref. \cite{csorgo99}], provided 
that particles are chaotically produced; $\lambda = f_c^2 (= (\langle n_{core} 
\rangle/\langle n_{tot}\rangle)^2)$. Moreover, if we can assume the 
laser-optical approach for the cross mark ({\large $\times$}) in Fig. 
\ref{fig2}, we obtain Eqs. (\ref{eq12}) $\sim$ (\ref{eq15}). By making use of 
them, we obtain Table \ref{table3} and Fig. \ref{fig5}. There is common region 
among results from analyses of the BEC of $3\pi$ and $3\pi\to 2\pi$ processes, 
$0.6 \lesim p \lesim 1.0$ and $f_c \sim 0.7$. However, the magnitude of the 
interaction regions estimated by the BEC of $3\pi$ and $3\pi\to 2\pi$ process 
are different. This problem should be considered in the future. 

It should be noticed that also recent data on the BEC of $3\pi^-$ reported by 
OPAL Collaboration \cite{opal98} suggest the necessity of introduction of some 
degree of coherence or $f_c$ in Ref. \cite{csorgo99}
\footnote{
Analysis of these data will be reported elsewhere \cite{biya00}. 
}. 

%
\section*{Acknowledgements}
Authors would like to thank Y. Nambu, S. Oryu and E. O. Alt for their kind 
suggestions and useful information. They are also indebted to G. Wilk for 
reading the manuscript. Our numerical calculations were partially carried out 
at RCNP of Osaka University. One of author (M. B.) is partially indebted to 
Japanese Grant-in-Aid for Education, Science, Sports and Culture 
(No. 09440103).

%

%

\section*{Addendum}
Because  we had no sufficient information on the raw data and 
corrected data in 2001,  we utilized the corrected data with 
$ K_{\rm Coulomb} $  in our analyses.  Of course we mentioned that 
our Eqs.~(9), (12), and (14) are available for the corrected  data including  
$K_{\rm SPC} $ (single particle correction (SPC)) and $K_{\rm acceptance}$, 
explicitly.
 Referring to Refs.~\cite{Schmidt-Sorensen:nd,Adams:2003vd}, we have examined the methods of 
correction used by NA44 Collaboration [17].   
In this addendum, thus we can present the data including $K_{\rm SPS}$ and 
$K_{\rm acceptance}$ and analyze them by means of Eqs.(9), (12) and (14). Before
 concrete analyses, we categorize raw and corrected data in Table~\ref{table4}.
%
%
\begin{table}[htb]
  \centering
  \caption{Category of data. Notice that $R =$ input is necessary for $K_{\rm Coulomb}$, where $R = 5$ fm is used in Refs.~[17,25,26].}
  \label{table4}
  \begin{tabular}{ll}
  \hline
  1) Raw data & raw data\\
  2) Quasi-corrected data (Q-CD) & $({\rm raw\ data}) \times K_{\rm SPC} \times K_{\rm acceptance}$\\
  3) Corrected data with Gamow & $({\rm raw\ data}) \times K_{\rm SPC} \times K_{\rm acceptance} \times K_{\rm Gamow}$\\
  4) Corrected data with Coulomb & $({\rm raw\ data}) \times K_{\rm SPC} \times K_{\rm acceptance} \times K_{\rm Coulomb}$ \\
  \hline
  \end{tabular}
\end{table}

\begin{figure}[htb]
  \centering
  \epsfig{file=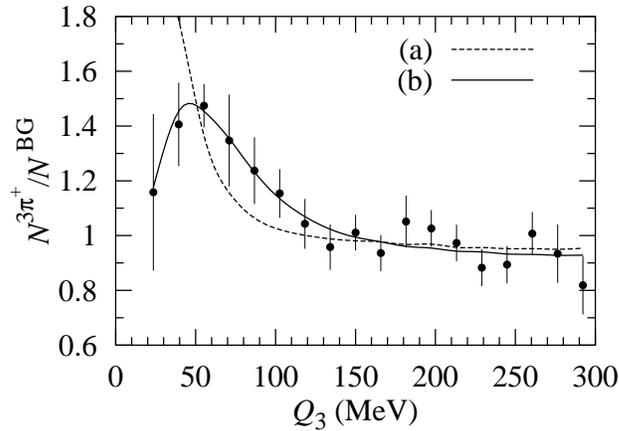,height=60mm}
  \caption{Analyses of quasi-corrected data (Q-CD). (a) and 
  (b) are results of Eqs. (5) and (9), respectively. .}
  \label{fig6}
\end{figure}

Using the data in Ref.~[17] with $K_{\rm Coulomb}$ or $K_{\rm Gamow}$, 
we can obtain the quasi-corrected data (Q-CD) shown in Fig.~\ref{fig6}, which are 
available for our purpose.  Using Eqs.~(5) and (9), we obtain the results shown in Table 5.  As seen in Fig. 6, the 
our formulation including the degree of coherence $\lambda$, i.e., Eq.~(9), 
seems to be available for analyses of the charged $\pi$ BEC.

%
%
%
%
\begin{table}[htb]
  \centering
  \caption{Estimated values for the data [17] by Eqs. (5) and (9) using CERN-MINUIT program.}
  \label{table5}
  \begin{tabular}{ccccc}
  \hline
  Formulas & $C$ & $R$ [fm] & $\lambda$ &$\chi^2/N_{dof}$\\
  \hline
  Eq. (5) & 0.94$\pm$0.03 & 5.83$\pm$0.40 & --- & 30/16\\
  Eq. (9) & 0.91$\pm$0.03 & 2.89$\pm$0.39 & 0.45$\pm$0.05 & 6.7/15\\
  \hline
  \end{tabular}
\end{table}

\begin{figure}[htb]
  \centering
  \epsfig{file=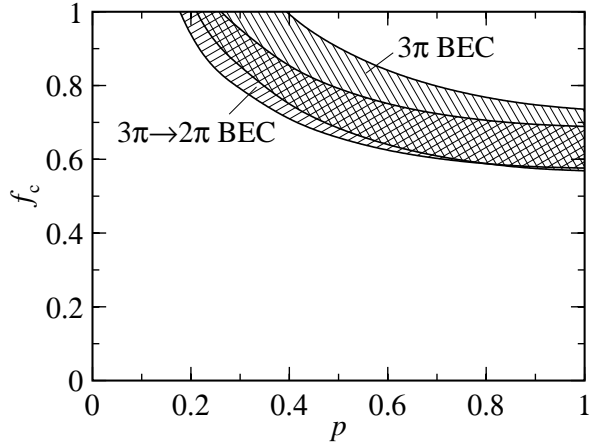,height=60mm}
  \caption{Sets of $f_c$ and $p$ estimated in analyses of BEC of two-charged 
  and three-charged pions using the CERN-MINUIT program. Widths of $f_c$'s 
  stand for error bar of $\pm 2\sigma$.}
  \label{fig7}
\end{figure}

To examine availability of the core-halo model described by the 
Coulomb wave functions, i.e., Eqs.~(12) and (14), we analyze the 
quasi-corrected data (Q-CD).  Our results are shown in Fig.~\ref{fig7} and Table~\ref{table6}.  
Coincidences between $3\pi$ BEC and $3\pi \to 2\pi$ BEC in Fig.~\ref{fig7} 
( $f_c$ vs. $p$) seem to be more enlarged than those of Fig.~5.

In this addendum, we have analyzed the quasi-corrected data (Q-CD) named in 
Table~4, utilizing Eqs.~(5), (6), (12) and (14). It can be said that 
our formulation works well for the description of the charged $3\pi$ BEC.

%
%
%
%
\begin{table}[htb]
  \centering
  \caption{Typical results from analyses of data of NA44 Collaboration by Eqs.~
 (12)$\sim$(15). $R_{\rm plane} = (3/2)R$, where $R_{\rm plane}$ is corresponding to a parameter in the plane wave formulation.}
  \label{table6}
  \begin{tabular}{cccc}
  \hline
  $p$ & 1.0 & 0.8 & 0.6\\
  \hline
  \multicolumn{4}{c}{$3\pi$ BEC}\\
  $f_c$                & 0.67$\pm$0.04 & 0.70$\pm$0.04 & 0.76$\pm$0.05\\
  $\lambda_3^*$        & 0.75          & 0.77          & 0.78\\
  $R$ (fm)             & 2.89$\pm$0.39 & 3.22$\pm$0.46 & 3.47$\pm$0.51\\
  $R_{\rm plane}$ (fm) & 4.33$\pm$0.58 & 4.83$\pm$0.69 & 5.21$\pm$0.76\\
  $\chi^2/N_{dof}$     & 6.7/15        & 6.8/15        & 6.8/15\\
  \multicolumn{4}{c}{$3\pi \to 2\pi$ BEC}\\
  $f_c$            & 0.67$\pm$0.03 & 0.69$\pm$0.03 & 0.73$\pm$0.03\\
  $\lambda_2^*$    & 0.45          & 0.46          & 0.45\\
  $R$ (fm)         & 4.42$\pm$0.38 & 5.01$\pm$0.45 & 5.49$\pm$0.50\\
  $\chi^2/N_{dof}$ & 14.5/17       & 14.7/17       & 14.7/17\\
  \hline
  \end{tabular}
\end{table}

As seen in the explanation above,  Fig.~1(a) and the upper line (Eq.~(5)) of 
Table 1 , and Fig.~3(a) and the upper line (Eq.~(9)) of Table 2 should be 
replaced by   Fig. 6 and  Table 5. Moreover,  Fig.~5 and Table~4 should be replaced by Fig.~\ref{fig7} and Table~\ref{table6}.

One of authors (M.~B.) would like to thank H. B\o ggild, T. Cs\"org\H o and B. L\"orstad for useful conversations.

%

%
\end{document}